\pdfoutput=1 
\documentclass[conference]{IEEEtran}
\usepackage{color}
\usepackage[pdftex]{graphicx}
\usepackage{array}
\usepackage{multirow}
\usepackage{hhline}
\usepackage{bm}
\usepackage{amsmath}
\usepackage{amssymb}
\usepackage{algorithm} 
\usepackage{algpseudocode}
\usepackage{amsmath}
\ifCLASSINFOpdf

   \usepackage{float}
   \usepackage{stfloats}
\usepackage{times}
\usepackage[para]{threeparttable}
\usepackage{array,booktabs,longtable,tabularx}
\newcolumntype{L}{>{\raggedright\arraybackslash}X}
\usepackage{ltablex}
\usepackage{siunitx}
\usepackage[flushleft]{threeparttablex}

\usepackage{etoolbox}
\makeatletter
\patchcmd{\@maketitle}
  {\addvspace{0.5\baselineskip}\egroup}
  {\addvspace{-1\baselineskip}\egroup}
  {}
  {}
\makeatother

   \graphicspath{{../pdf/}{../jpeg/}}

   \DeclareGraphicsExtensions{.eps,.pdf,.jpeg,.png}
\else

\fi

\DeclareRobustCommand*{\IEEEauthorrefmark}[1]{%
  \raisebox{0pt}[0pt][0pt]{\textsuperscript{\footnotesize\ensuremath{#1}}}}
  
\begin{document}
\title{PRUNIX: Non-Ideality Aware Convolutional Neural Network Pruning for Memristive Accelerators}
\author{
    \IEEEauthorblockN{Ali Al-shaarawy\IEEEauthorrefmark{1}, Amirali Amirsoleimani\IEEEauthorrefmark{2}, Roman Genov\IEEEauthorrefmark{1}}
    \IEEEauthorblockA{\IEEEauthorrefmark{1}Department of Electrical and Computer Engineering, University of Toronto, Toronto, Canada \\
    \IEEEauthorrefmark{2}Department of Electrical Engineering and Computer Science, York University, Toronto, Canada \\
    Email: ali.alshaarawy@mail.utoronto.ca,
    amirsol@yorku.ca,
    roman@eecg.utoronto.ca }
}
\markboth{IEEE TRANSACTIONS ON NEURAL NETWORKS AND LEARNING SYSTEMS,~Vol.XX, No.XX, June~2020}
{Shell \MakeLowercase{\textit{et al.}}: Bare Demo of IEEEtran.cls for IEEE Journals}

\maketitle

\begin{abstract}
 In this work, PRUNIX, a framework for training and pruning convolutional neural networks is proposed for deployment on memristor crossbar based accelerators. PRUNIX takes into account the numerous non-ideal effects of memristor crossbars including weight quantization, state-drift, aging and stuck-at-faults. PRUNIX utilises a novel Group Sawtooth Regularization intended to improve non-ideality tolerance as well as sparsity, and a novel Adaptive Pruning Algorithm (APA) intended to minimise accuracy loss by considering the sensitivity of different layers of a CNN to pruning. We compare our regularization and pruning methods with other standards on multiple CNN architectures, and observe an improvement of 13\% test accuracy when quantization and other non-ideal effects are accounted for with an overall sparsity of 85\%, which is similar to other methods.
 
\end{abstract}

\vspace{-0.3cm}
\IEEEpeerreviewmaketitle

\section{INTRODUCTION}

\IEEEPARstart{M}{emristor} devices have been receiving significant attention due to their ability to perform vector matrix multiplication (VMM) in the analog domain [1], which can be exploited to accelerate the operation of deep neural networks (DNNs). These systems have been implemented using a wide range  of  emerging  technologies  such  as, resistive  random access memory (ReRAM) [2], phase change memory (PCM) [3] and magneto-resistive RAM (MRAM) [4]. Despite  their  many  advantages,  the  analog  nature  of  computation  using these devices  poses  certain  challenges owing  to  device-  and  circuit-level  non-idealities  such as,  interconnect  parasitics,  process  variations  in  the  synaptic devices, driver and sensing resistances, etc. [5]. These non-idealities result in accuracy degradation of the programmed conductance values, leading to errors in the VMM calculations, which lowers the overall performance of the DNN. Another challenge faced when implementing DNNs on a crossbar architecture stems from the fact that state of the art DNNs are extremely over-parametrized, thus requiring a massive amount of resources to operate. One solution to this would be pruning neural network layers to shrink the model size [6]. Crossbar-aware pruning has been explored in [7-9], but none so far have been able to account for all major crossbar non-idealities which alter the accuracy and stability of stored network parameters in one framework.

\section{PRELIMINARIES}

\subsection{Memristor Crossbars as Neural Network Accelerators}
One of the major properties of memristors is that their conductances are programmable. This feature, when coupled with a crossbar architecture, affords us the computation of vector-matrix multiplication in the analog domain [10], since VMM is the backbone of neural network computation, one can easily see where the interest in this technology by the deep learning community arose from. The VMM operation is achieved, as in Fig. 1, by applying voltages, to the rows of the crossbar and this generates a current in each bit-line $I_j = \sum_{i = 1}^{H} V_iG_{ij}$ where $G_{ij}$ is the programmed conductance value, representing the network weight. Since it is not possible to program memristor cells to negative conductances, two adjacent crossbars or bit-lines are used in tandem instead [11], such that the actual weight value is $G_{crossbar1} - G_{crossbar2}$, this also allows us to double the effective range of an $n$-bit quantization scheme without adding a complement bit, since both negative and positive weights can be represented in $n$-bits just on different crossbars, we will be utilising this fact throughout this work.

\subsection{Neural Network Pruning}
Pruning of neural networks is a method of reducing the computational and memory requirements for neural networks by systematically zeroing and removing parameters from an existing network [6].  Typically, the initial network is large and accurate, and  the  goal is  to  produce  a  smaller  network  with  similar accuracy. Pruning techniques can be categorized as either non-structured or structured. Non-structured pruning can achieve higher element-wise sparsity, the number of zeros in the network, but its irregularity makes it difficult to exploit in order to achieve any significant acceleration. Structured pruning, however, works by pruning entire structures (e.g. filters, matrix rows or columns) from the network, drastically reducing the computational resource consumption of a network.

\subsection{Crossbar Non-Idealities}

Hardware non-idealities are mainly products of the imperfect fabrication process of the crossbar, limitations of control circuitry, and device features of the memory cells themselves. They affect the accuracy, precision and longevity of crossbar mapped neural networks, so it is in our best interest to mitigate these effects as best as possible.
\subsubsection{Quantization}
Due to the limited resolution of the CMOS circuitry that is used to program the conductances of the crossbar, analog weights must be mapped to the nearest value represented by the discrete conductance states of the memristors $|w_l| \in [0, a_1, a_2,... a_{2^n-1}]$ Where $a_i$ are the values representable in an n-bit encoding scheme. This limits both the range and accuracy of network parameters, and can be detrimental to the crossbar's performance.
\subsubsection{Memristance  drift}
Another consequence of repeated programming of memristor devices is memristance drift; the deviation of memristor conductances from the programmed value. This deviation will accumulate as the device is repeatedly read unless it is reprogrammed after each read operation, and has been shown to propagate in deeper layers of a neural network where its negative impact on accuracy is amplified [12].
\subsubsection{Conductance range degradation}
During the programming process of the crossbar, conductance values are adjusted by the application of a relatively high voltage pulse to the memristor [16]. This high voltage may change the structure of the filament which has been known to degrade the usable conductance range of the device, and thus limits the range of network parameters which can be programmed onto the crossbar. Like drift, this effect also accumulates through repeated Programming.
\subsubsection{Stuck-at-faults}
A stuck-at-fault denotes a memristor with the conductance state fixated to a high (stuck-on) or low (stuck-off) conductance value. They can be products of both the fabrication process or heavy utilisation.

\begin{figure}[t]
\centering
\includegraphics[width=0.5\textwidth]{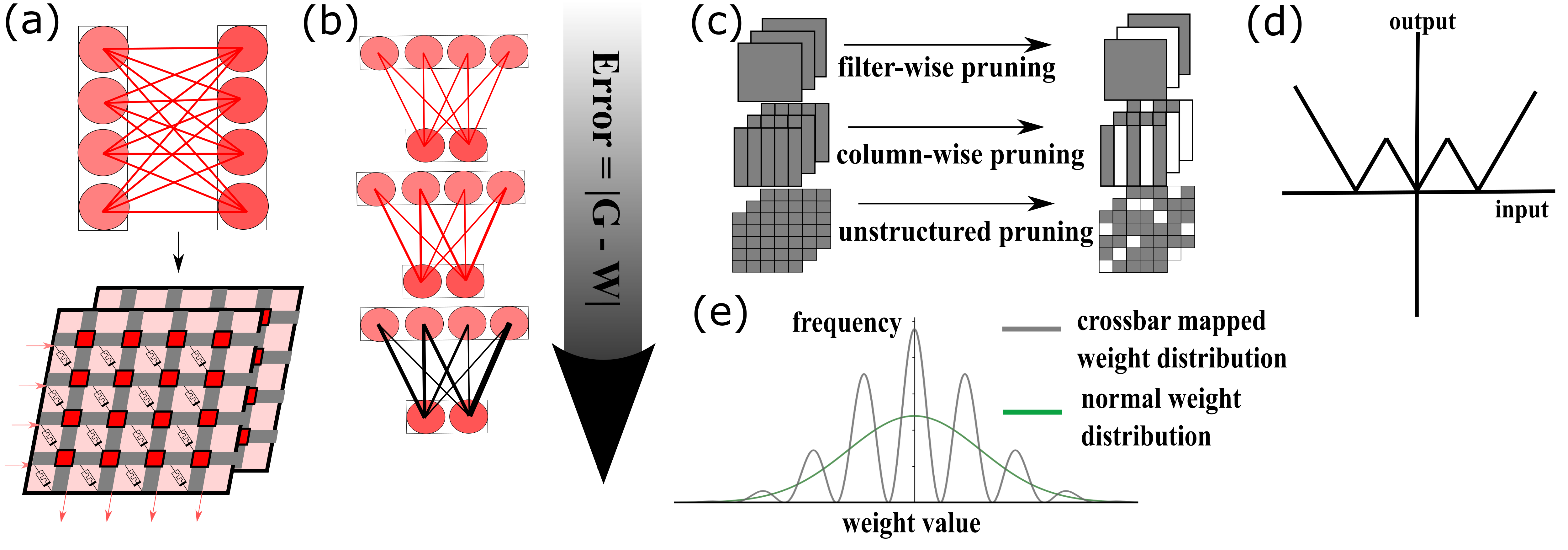}
\caption{ Illustration of crossbar mapping, pruning, error propagation and sawtooth regularization. (a) Neural network mapping onto crossbar. (b) Error of crossbar network increases with successive read and write cycles due to aging and state drift. (c) Different types of pruning. (d) Sawtooth regularization. (e) Weight distributions of mapped and unmapped NN weights.}
\end{figure}

\section{NON-IDEALITY AWARE PRUNING}
 In this section we present our CNN training framework which includes the Group Sawtooth Regularization scheme and Adaptive Pruning Algorithm, as well as an overall strategy to train networks in order to achieve an exceptionally high test accuracy while minimising accuracy loss when non-idealities are applied. Our method differs from existing pruned crossbar based accelerators [7-9] since we do not require a priori knowledge of our target platform architecture or any dedicated auxiliary circuitry, while accounting for accuracy loss due to quantization of weights, state drift, aging and stuck-off faults.
\subsection{Initial Training}
Network weights usually follow a quasi-normal weight distribution after training. Previous work has demonstrated that after quantization and mapping, the weight distribution will follow a more sinusoidal weight distribution with high weight density centered around the quantized weight values and tapering occurring due to fluctuations caused by state drift [13], as demonstrated in figure 1. Crossbar aware training through the use of sawtooth shaped norm regularisation [14],
\begin{equation}
\Gamma(W_l, p, a) = \bigg|clamp(\frac{W_l}{p},a) - \bigg \lfloor{\frac{W_l}{p} + \frac{1}{2}}\bigg \rfloor \bigg|
\end{equation} where $W_l$ is the weight tensor at a layer $l$, $clamp(x,a)=max(-a,min(x,a))\in [-a,a]$, $a$ is the maximum conductance level expressable, and $p$ is the periodicity of the quantization levels, has shown great promise in alleviating the effects of the discrepancy between trained and mapped weight distributions, as weights are pushed towards the minim during training, which align with the crossbar's conductances. We propose two modifications to this method in order to adapt it to our objectives of creating structured sparsity and reducing the effects of aging: we apply sawtooth norm over groups of weights to create greater structured sparsity similar to the L1 Group Lasso proposed by Wen {\it et al.} [15], Suppose that the weight matrix of a layer $l$ of our model can be expressed as 
\begin{equation}
  W_l \in
  \begin{cases}
                                   \mathbb{R}^{N_l \times C_l \times K_l \times K_l} & \text{if $l$ is Conv} \\
                                   \mathbb{R}^{N_l \times C_l} & \text{if $l$ is FC} \\
  \end{cases}
\end{equation} 
where $N_l$ , $C_l$ and $K_l$ represent input, output and kernel dimensions respectively. Our loss function can be expressed as 
\begin{equation}
L'(W_l) = L(W_l) + \lambda_s \sum_{l=1}^{L_f} \sum_{g=1}^{G} \Gamma(W_{lg}, p, a)\\
\end{equation} 
where $\mathrm{L}$ and $G$ represent cross entropy loss and the number of groups in a convolutional layer respectively and $\lambda_s$ is the group sawtooth coefficient. To mitigate the effects of the loss of conductance levels, we limit $a$ such that it is less than the maximum programmable conductance level. Our testing demonstrates that these modifications significantly improve performance in comparison to L1 Group Lasso and Sawtooth Regularization methods when a broad range of non-idealities are simulated.

\subsection{Pruning}

Since Convolutional layers constitute a majority of the $\mathrm{FLOP}$s of the feed forward step in a CNN, our method will focus on them. Convolutions are unrolled from 
\begin{equation} 
\begin{gathered} 
\begin{bmatrix}
    a_{11} & a_{12} & \dots \\
    \vdots & \ddots & \\
    a_{M1} &        & a_{MN} 
\end{bmatrix}
\circledast
\begin{bmatrix}
    k_{11} & k_{12}  \\
    k_{21} & k_{22} 
\end{bmatrix} 
to \\
\begin{bmatrix}
    a_{11}&a_{12}&a_{13}&\dots &a_{MN}
\end{bmatrix} 
\times
\begin{bmatrix}
    k_{11} & 0 & \dots&0\\
    k_{12} & k_{11} &\dots&0\\
    \vdots & \vdots &\ddots & \\
    k_{21} & 0&\dots&0\\
    k_{22}&k_{21}&\dots&0\\
    \vdots & \vdots &\ddots & \\
    0 & 0 & \dots&k_{22}
\end{bmatrix} 
\end{gathered} 
\end{equation} 
\begin{figure*}[t]
\centering
\includegraphics[width=\textwidth]{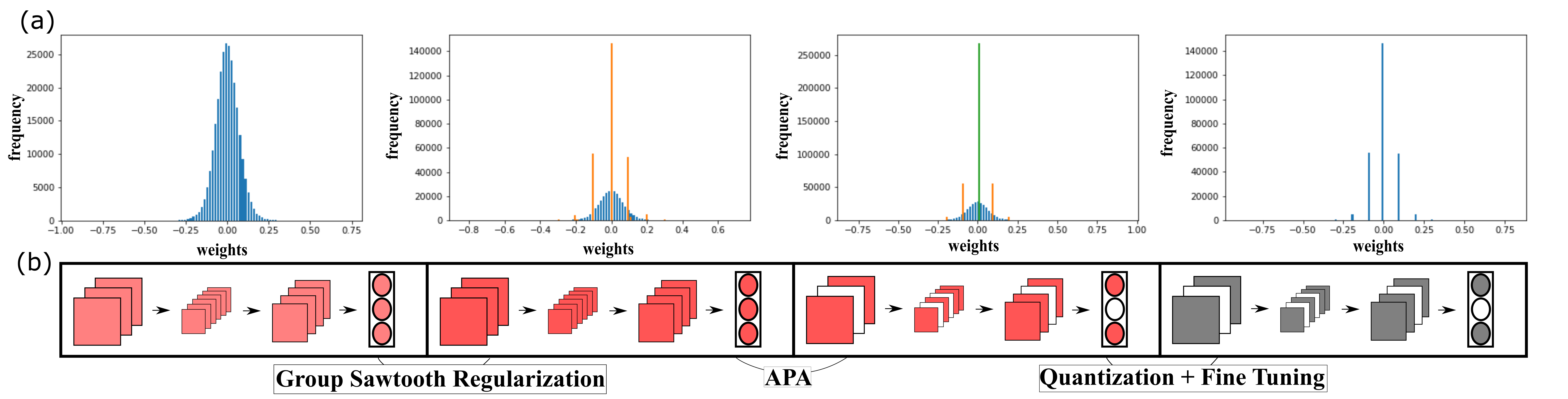}
\caption{ Overview of stages in PRUNIX and weight distributions of the resultant network after each stage. (a) Weight distributions after each stage of PRUNIX for 4-bit quantization. (b) High-level overview of training stages: initialisation, regularised training, pruning and quantization and fine-tuning. }
\end{figure*}

in order to convert them to the VMM domain. The input and kernel have initial dimensions $M\times N$ and $k\times k$ respectively. The equivalent input vector has dimensions $1\times(M*N)$ and the equivalent kernel matrix has dimensions $(M*N)\times(\frac{M-k+2P}{S_h}+1)(\frac{N-k+2P}{S_w}+1)$, where $P, S_h$ and $S_w$ are the input padding and stride dimensions respectively. The vertical distance between $k_{12}$ and $k_{21}$ is $M-k$.
From this, we can see that pruning columns or rows from filters will have little effect on compression, as we will not be able to remove any columns or rows from our equivalent kernel matrix. We also consider that unstructured sparsity can be used to decrease the effect of stuck-off faults by increasing the likelihood that affected crossbar cells overlap with sparse weights [17]. One final consideration is the effects of pruning on different layers [6], as they extract different features from the input image. This means that our pruning algorithm must consider the effects of each layer on accuracy before pruning. The APA zeros out a percentage of weights and filters with the lowest magnitude, adjusting this percentage at each layer to minimise accuracy loss. The tunable parameters offered by the APA allow us to better optimise filter-shaped and unstructured sparsity to achieve high compression and fault tolerance, this differs from existing methods which focus on only one type of sparsification, and do not adjust pruning across different layers

\begin{table}[t]
\renewcommand{\arraystretch}{1.15}
\centering
\caption{ Adaptive Pruning Algorithm parameter definitions}\label{comparison_SOTA_detection}
\begin{tabular}{lcrcrr}
\toprule \toprule
\textbf{Parameter} & \textbf{Definition}\\ 
 $\lambda_p$ &  initial pruning rate\\ 
 $\mu$ &  fraction of pruning rate for unstructured pruning\\
 $\sigma$ & minimum accuracy loss at layer \\
 $\gamma$ & pruning rate decrease factor \\ [1ex] 
\bottomrule \bottomrule
\end{tabular}

\end{table}

\begin{algorithm}[t]
	\caption{Adaptive Pruning Algorithhm} 
	 \hspace*{\algorithmicindent} \textbf{Input: $model$, $\lambda_p$, $\mu$, $\sigma$, $\gamma$} \\
     \hspace*{\algorithmicindent} \textbf{Output: $pruned$ $model$} 
	\begin{algorithmic}[1]
		\For {$l$ in model}
		\While {accuracy loss $> \sigma$}
		    \If{$l$ is Conv.}
    				\State UnstructurePrune(l, $\lambda \times \mu$
    				\State FilterPrune(l, $\lambda \times (1-\mu)$)
    		\ElsIf{$l$ is FC.}
    			\State UnstructurePrune(l, $\lambda \times \mu$)
    		\EndIf
    		\If{accuracy loss $> \sigma$}
    		    \State undo pruning
    		    \State $\lambda \gets \lambda \times \gamma$ 
    		\EndIf
    	\EndWhile
		\EndFor
	\end{algorithmic} 
	
\end{algorithm}
\subsection{Quantization and Fine-tuning}
The final stage of our training framework involves fully quantizing our network and relaxing regularization. relaxing regularization has been shown to recover lost accuracy while maintaining the desirable characteristics obtained from previously applied regularization [18]. We demonstrate that this holds true for Group Sawtooth Regularization as well. We relax regularization by decaying $\lambda_s$ by a constant factor each epoch to reduce the impact of Group Sawtooth Regularization on weight gradients.

\begin{figure*}[t]
\centering
\includegraphics[width=\textwidth]{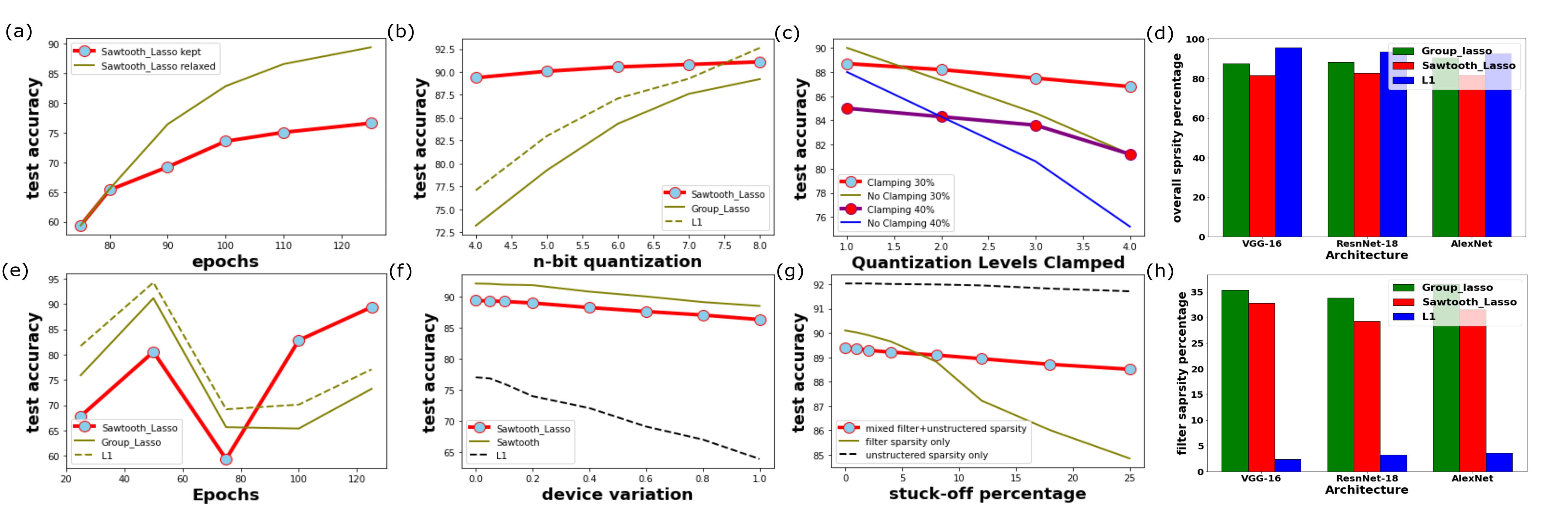}
\caption{ Test accuracies when non-idealities are considered, and sparsities. (a) Comparison between keeping and relaxing regularization at final stage of training on ResNet-18. (b) Comparison of test accuracy with different quantization bit lengths on ResNet-18. (c) Comparison aging on models trained with and without a clamped quantization range. (d) Comparison of overall sparsity. (e)  Comparison of test accuracy through training on ResNet-18 (f)  Comparison of test accuracy against different levels of state drift acting on 30\% of cells on ResNet-18. (g) Comparison of test accuracy against percentage of stuck-off cells on ResNet-18. (h) Comparison of percentage of sparse filter under same test conditions.}
\vspace{-4mm}
\end{figure*}

\begin{figure}[t]
\centering
\includegraphics[scale = 0.25]{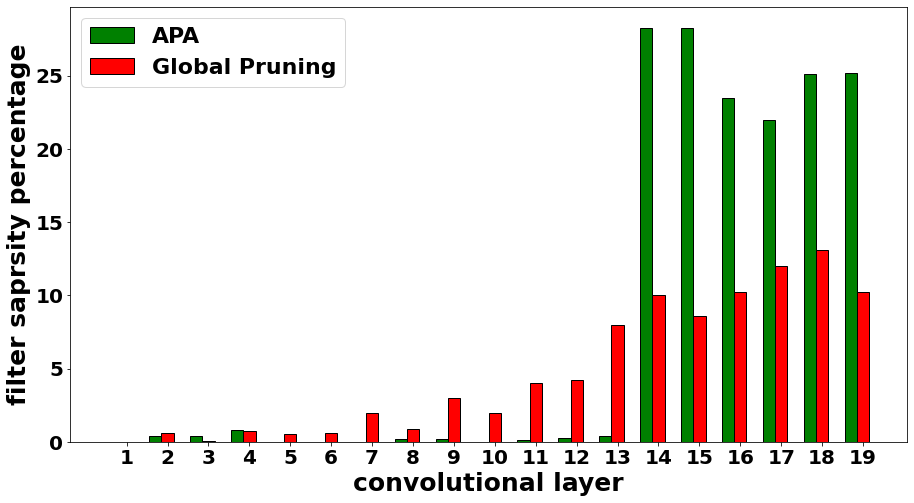}
\caption{ Comparison of Filter sparsity across all convolutional layers of ResNet-18 between global pruning and APA.}
\vspace{-4mm}
\end{figure}

\section{EXPERIMENTAL RESULTS}
In this section, we will be evaluating our training framework on the CIFAR-10 data set using the VGG-16, ResNet-18 and AlexNet CNN architectures. PRUNIX model construction, training and non-ideality simulation are performed using the Pytorch API. We first compared the performance of our group sawtooth regularization to group lasso and L1 regularization in three tests: test accuracy with n-bit quantization, overall sparsity percentage and sparse filter percentage. PRUNIX was able to secure a 13\% greater accuracy than both with 4-bit quantization, with stable accuracy on varying bit lengths of quantization. Our method also achieved similar overall sparsity as the other two regularziation method at 80\% to 90\%, with filter shaped sparsity similar to group lasso at 30\%, it should be noted that all these tests were conducted using the APA, with the same parameters. 

We then tested the performance of our method when device faults are taken into account. We began by testing how effective the combination of filter and unstructured sparsity was in mitigating the effects of stuck-off faults. Stuck-off faults were modelled by setting a varying percentage of our weights to 0, applied with equal probability across all weights, simulating stuck-off devices in a physical crossbar. Our method performed significantly better than purely filter shaped sparsity, and only marginally worse than pure unstructured sparsity in this test. We then measured the effectiveness of Group Sawtooth Regularisation against device variations, which were simulated by applying a random variation parameter sampled from a range, $\pm(\delta = random(0,r\hat{W_l}))$, where $r$ is the varying parameter and $\hat{W_l}$ is the mean weight value, across all weights of a 4-bit quantized model. The sampling distribution we used had a mean $\mu = 0.2r\hat{W_l}$, and standard deviation $\sigma = 0.1r\hat{W_l}$. Our method's performance in this test was comparable to pure sawtooth-regularization, suggesting that its acting on groups did not alter performance greatly. Finally, our performance against aging's effects was also measured, we limited the maximum conductance level of 30\% and 40\% of our weights by a varying amount, and compared the test accuracy with and without the limiting of the clamping range in our group sawtooth by 4 levels in both the positive and negative ranges of a 8-bit quantized model, as expected, this caused fewer weights to lie in the higher conductance levels, and as such, lower accuracy loss as those levels were lost to aging. We also observed a significant improvement in test accuracy if regularization is removed at the final stage of training. We also observed that the APA was better able to avoid pruning sensitive layers, namely deeper layers, than global pruning as demonstrated by Fig 4.

\section{Conclusion}
In this paper, we proposed PRUNIX: a software training framework which combines non-ideality aware regularization and pruning in order to train CNNs which are resilient to quantization of weights, loss of quantization levels, stuck-off faults and state drift. Our experimental results show a significant gain in accuracy when non-idealities are present in comparison to standard methods, while retaining the compression achievable with dedicated methods.

\end{document}